# Time-dependent photoionization of azulene. Competition between ionization and relaxation in highly excited states

Valérie Blanchet, Kevin Raffael, Giorgio Turri, Béatrice Chatel, Bertrand Girard

Laboratoire Collisions Agrégats Réactivité (UMR 5589, CNRS -Université de Toulouse, UPS), Institut de Recherches sur les Systèmes Atomiques et Moléculaires Complexes, France

and

Ivan Anton Garcia, Iain Wilkinson , Benjamin J Whitaker

School of Chemistry, University of Leeds, Leeds, LS2 9JT, United Kingdom

Correspondence should be addressed to V.B. (e-mail: val@irsamc.ups-tlse.fr)

ABSTRACT

Pump-probe photoionization has been used to map the relaxation processes taking place from highly vibrationally excited levels of the $S_2$ state of azulene, populated directly or via internal conversion from the $S_4$ state. Photoelectron spectra obtained by 1+2' two-color time-resolved photoelectron imaging are invariant (apart from in intensity) to the pump-probe time delay and to pump wavelength. This reveals a photoionization process which is driven by an unstable electronic state (e.g. doubly excited state) lying below the ionization potential. This state is postulated to be populated by a probe transition from $S_2$ and to rapidly relax via an Auger like process onto highly vibrationally excited Rydberg states. This accounts for the time invariance of the photoelectron spectrum. The intensity of the photoelectron spectrum is proportional to the population in $S_2$. An exponential energy gap law is used to describe the internal conversion rate from $S_2$ to $S_0$. The vibronic coupling strength is found to be larger than 60±5 μeV.

## I. INTRODUCTION

The electronic structure of azulene is noteworthy due to the atypical fluorescence which occurs from the second excited $S_2(2A_1)$ state, instead of the lower $S_1(1B_2)$ excited state. As such, it is well known as the textbook exception to Kasha's rule that "the emitting level of a given multiplicity is the lowest excited level of that multiplicity".[1] Little is known about the higher electronic valence states,[2-4] other than their use as stepping-stones through which to prepare highly excited molecules in order to observe the unimolecular dynamics of the molecule.[5,6]

Azulene is a planar asymmetric top molecule that belongs to the $C_{2v}$ point group (see Figure 1). There are four distinct regions in the absorption spectrum of the molecule; 700-500 nm, 350-310 nm, 290-260 nm, and 240-200 nm with relative oscillator strengths 0.01, 0.06, 1, and 0.4 (see Table 1). The $S_2(2A_1) \leftarrow S_0(1A_1)$ and the $S_4(3A_1) \leftarrow S_0(1A_1)$ absorptions are optically allowed with an electric dipole transition moment along the $z$ axis. The $S_1(1B_2) \leftarrow S_0(1A_1)$ and $S_3(2B_2) \leftarrow S_0(1A_1)$ transitions are $y$ polarized. Like the $S_1(1B_2) \leftarrow S_0(1A_1)$ transition the $S_2(2A_1) \leftarrow S_0(1A_1)$ transition is characterized by the creation of double bond type character on the transannular connection. Early fluorescence quantum yield experiments suggested long lifetimes[7] for single vibrational levels in $S_2$ and sub-picosecond ones for the $S_3$ and $S_4$ levels.[2] The 2-3 ns lifetimes of the $S_2$ vibrational levels were later confirmed by picosecond-resolved quantum beat experiments.[8,9] Intriguingly, Diau *et al.*[10] observed an additional fast exponential component in the ion time-profile in two colour photoionization experiments, with a 350 fs decay on the top of the long decay component when exciting the $S_2$ state with 470 meV excess vibrational energy. We have performed time resolved pump-probe photoionization and photoelectron spectroscopy via the electronically excited states ($S_2$ to $S_4$) of azulene. At all the wavelengths studied here the transient signals exhibit two distinct and well-defined behaviours: (i) Short-term (on the order of a picosecond) polarization dependent transients[11] and (ii) longer (10 ps – 1 ns) time-scale decays. The present paper will consider the longer term behaviour, which will be explained as the result of internal conversion from $S_2$ to $S_0$ and treated via a statistical model, and focus principally on the photoionization processes occuring within the electronically excited states of azulene.

Although for most of our experiments a 1+1' photoionization process is energetically allowed, we will present a series of converging arguments to conclude that the photoionization actually proceeds via a complex (1+2') mechanism with fast internal conversion taking place within the probe pulse. To elucidate this photoionization mechanism we have measured pump-probe (263 - 295 nm / 400 nm) signals on the parent photoion and also recorded photoelectron spectra; with both one-colour (258 and 266 nm) and two colour (268 - 335 nm / 400 nm) ionization schemes. Although the ionization mechanism is indubitably complicated, we will present evidence suggesting that the phenomena we are observing may be quite generic in polycyclic molecules.

Figure 1 provides a framework for the forthcoming discussion. It depicts a number of possible excitation and ionization paths. Fig 1.1 is a schematic of the one-colour two photon ionization process and Fig 1.2 and 1.3 are those of the two-colour photoionization processes. Each electronically excited state can be either directly excited/ionized (paths (a) and (c)) or each can first decay to a lower electronic state by vibronic coupling before being ionized (paths (b) and (d-e)). In this latter case two probe photons are necessary to ionize the molecule due to the large amount of vibrational energy in the lower lying states. In the two-colour photoionization schemes, at least one (resonant or non-resonant) intermediate state mediates the transition. In the resonant case, this intermediate state can be directly ionized (path (d)) or it can relax towards another electronic state if its lifetime is shorter than the probe pulse duration (path (e)).

In this paper we will examine successively: Koopmans' type correlations in the $S_2$ and $S_4$ electronic states[12] via the one-colour photoelectron spectrum, the evidence for the presence of a doubly excited electronic state involved in the photoionization via the two-colour time-resolved photoelectron spectroscopy and its relaxation onto Rydberg states. Finally we complete this study with a statistical approach to describe the long decay component and the coupling between the $S_2$ and $S_0$ states.

## II EXPERIMENTAL DETAILS

Experiments were performed in our two laboratories using similar instruments. We describe here the experiment in Toulouse in detail and then point out any significant differences between the two set-ups. We employ pump-probe time-of-flight (TOF) mass-spectrometry to detect the photoions and velocity map imaging to record the photoelectrons spectra.[13]

Both pump and probe pulses were generated from a master 1 kHz 2.5 mJ regenerative amplifier centred at ~805 nm and delivering a pulse with a Fourier limited Full-Width at Half Maximum (FWHM) of ~60 fs (Amplitude Systems). A home-made non-collinear optical parametric amplifier (NOPA),[14] pumped with a fraction of the frequency doubled output of the Ti:sapphire amplifier, with subsequent second harmonic generation provided the ultraviolet (UV) pump pulse. The NOPA output was partly recompressed by a prism pair. The typical FWHM of the pump pulse in energy was 30-50 meV corresponding to a Fourier limit of 35-60 fs. For most of the experiments reported here the probe pulse, at ~403 nm, was obtained by second harmonic generation (SHG) of the fundamental output of the regenerative amplifier. The cross-correlation between the pump and probe pulses, recorded by off-resonant multiphoton ionization of nitric oxide, had a typical 120 fs FWHM. Pump and probe beams were combined at a small angle (~1°) and focused onto the molecular beam by a 750 mm focal length spherical aluminium mirror. In order to reduce multiphoton effects, such as dissocitative ionization, the typical energies used were <1 μJ for the pump pulse and ~15 μJ for the probe pulse. The pulse-to-pulse stability was better than 85% for both pump and probe beams. The dimensions of the two beams were measured at their spatial overlap, from which we deduce intensities of ~3.0 $10^{10}$ W/cm$^2$ for the pump pulse and ~2.0 $10^{11}$ W/cm$^2$ for the probe pulse. These intensities correspond to an insignificant ponderomotive potential of a few meV. We are therefore confident that multiphoton ionization takes place in an unperturbed energy scheme.[15]

The set-up in Leeds is very similar except that the regenerative amplifier (Clark MXR 2010) is seeded with a frequency doubled Er fibre laser and consequently the central wavelengths of the harmonics that are used for the probe and pump fields are slightly shifted from those used in Toulouse: 387 and 258 nm (Leeds), 403 and 269 nm (Toulouse). In addition a Nd:YAG pumped dye laser (Continuum Surelite/Sirah) was used in the Leeds apparatus to obtain photoelectron images on the nanosecond time-scale. For this experiment the third harmonic of one Nd:YAG laser was used to pump a dye (Exalite 404) to generate tunable light close to 400 nm. The line width of the dye laser was 0.1 $cm^{-1}$ with a typical pulse energy kept below 3 mJ to avoid fragmentation. In all of the experiments reported below every signal has been recorded with parallel laser polarisation.

Azulene molecules (Aldrich, 99% without further purification) were sublimed continuously at 340 K with 150 Torr of nitrogen and then expanded through a 200 μm diameter nozzle. The resulting molecular beam was collimated by a 1 mm diameter skimmer before it intersected, perpendicularly, the focused laser beams. The resulting ions and photoelectrons were subsequently detected at the output of a 40 cm time-of-flight velocity map imaging spectrometer whose axis is perpendicular to the plane defined by the lasers and molecular beam.[13] The azulene sublimed cleanly without evidence of decomposition in the ion mass spectrum observed with either pump or probe laser alone. We were unable to detect any trace of van der Waals clusters such as azulene-$N_2$ or $(azulene)_n$ in the mass spectrum under these expansion conditions.[16,17]

For each pump-probe delay, the ion signal was averaged over ~2000 laser pulses and the photoelectron image over $\sim 4\times10^5$ laser pulses. The images were calibrated by recording photoelectron signals from nitric oxide, acetylene and oxygen at various accelerating and focussing voltages. Typically the detection window spanned 0.1 to 3.5 eV with a resolution of 65 meV at 1.4 eV (for a repeller voltage of -3kV and an extractor to repeller voltage ratio of 0.735.[13]). The 5% energy resolution achieved corresponds to the laser bandwidth.

## III RESULTS AND DISCUSSION

### A - Pump-Probe signals

Figure 2 shows the typical monoexponential decay of the parent $C_{10}H_8^+$ ion recorded for two pump wavelengths respectively falling within the $S_0$-$S_4$ ($\lambda_{pump}$=268 nm) and $S_0$-$S_3$ ($\lambda_{pump}$=293 nm) absorption bands. A monoexponential picosecond decay has been observed for all the pump wavelengths investigated between 258 and 293 nm. In these experiments the probe pulse is the second harmonic of the laser chain, namely 403 nm, for all pump wavelengths except 258 nm where the probe wavelength was 387 nm. This detection scheme corresponds to paths (c-e) in Figure 1.

The pump photon energy in these data spans the absorption range of the $S_3$ and $S_4$ electronic states (see Table 1),[2] but is always below the photodissociation threshold[18] or the isomerization barrier to naphtalene.[5,19] The decay time decreases with increasing excitation energy but is greater than 30 ps even for the highest energy studied, ~5 eV. These would be remarkably long lifetimes for such high lying electronic states and would not be consistent with the broad features observed in the jet-cooled fluorescence excitation spectra of the $S_3$ and $S_4$ states.[2] From the spectral profiles Fujii *et al.*[2] suggested that the $S_3$ and $S_4$ states were efficiently coupled to the $S_2$ state and this was later confirmed by Lawrance and Knight by pump-probe spectroscopy.[20]

In order to assign the non-radiative processes related to the decay times shown in Figure 2 and, in particular, to evaluate the relative contributions from paths (c) and (d) to the observed ion transients we investigated photoelectron spectra, as described below.

**B - One-colour photoelectron spectra**

The signature of the $S_4$ or $S_3$ states via photoionization has never been detected. Indeed, the 1+1 photoelectron spectrum recorded at the origin of the $S_4$ band with an 8 ps pulse results in a single photoelectron peak from the $S_2$ electronic state, and is correlated to the cation ground state via a $\Delta v \approx 0$ propensity rule.[21] This is indicative of efficient internal conversion from the initially pumped vibrationless level of the $S_4$ state onto $S_2$, as depicted by path (b) in Fig.1. A similar propensity to preserve the vibrational quantum state upon photoionization from $S_2$ was observed in the 1+1' zero kinetic energy (ZEKE) photoelectron spectra recorded on resonance via the first few vibrational levels of $S_2$.[17] These experiments show that, on the picosecond timescale, the prevailing contribution to the molecular eigenstates underlying the electronic absorption spectrum around the $S_4$ electronic origin comes from $S_2$. The same conclusion can be drawn from similar experiments performed close to the origin of the $S_3$ state.[21] On the picosecond timescale, no clear signature of photoionization to $D_1$ is observed even though this photoionization route is energetically allowed. The observation of a $\Delta v \approx 0$ propensity rule suggests that the $S_2$ and cation ground states should have a similar geometry.

In order to detect the $S_4$ signature via photoionization, we performed the same experiment but with 120 fs pulses. The photoelectron spectrum was recorded for two different wavelengths (258 nm and 266 nm) in which the total two-photon energy (9.61 and 9.32 eV respectively) was sufficient to photoionise azulene to the cationic ground state, $D_0$ (7.41 eV), or the first excited state of the ion, $D_1$ (8.50eV), but not to the second excited state $D_2$ (10.07 eV) (see Table 1).[17,22]

Figure 3 shows that there are at least two thresholds on the photoelectron spectrum at both wavelengths. The vibrational energy in $S_2$ is 1.24 eV at 258 nm and 1.10 eV at 266 nm. Assuming the same $\Delta v \approx 0$ propensity rule onto $D_0$ observed by Weber and Thantu,[21] the photoelectron spectra would be anticipated to peak at 0.96 eV and 0.81 eV for the two excitation wavelengths respectively. But, as can be seen in Figure 3, there are no obvious discrete peaks appearing at these energies. Instead, the large rising features around 1.11 and 0.82 eV fit exactly the ionization threshold onto the $D_1$ ion state. It seems that on the timescale of a few tens of femtoseconds ionization to $D_1$ is favoured, in marked contrast to what is observed on the picosecond timescale. The vibrational energy in the $S_4$ state is respectively 0.26 eV at 266 nm and 0.4 eV for a pulse centred at 258 nm.[2] For 266 nm excitation, the photoelectron spectrum peaks at 0.55 eV, corresponding to a vibrational energy in $D_1$ of 0.27 eV. This agrees perfectly with photoionization onto $D_1$ governed by a $\Delta v \approx 0$ propensity rule, but this time occurring out of the $S_4$ electronic state. The same $\Delta v \approx 0$ propensity rule applied at 258 nm should lead to a photoelectron spectrum peaking at 0.71 eV, at which we observe the onset of a plateau. The fraction observed in the $D_0$ channel (lying from 1.9 to 0.9 eV on Fig. 3 for the 266 nm photoionization case) is formed with very little vibrational selectivity; the tail spreads over more than 1 eV. This can be tentatively assigned to a growing contribution of the $S_2$ electronic state in the one-photon resonance over the 120 fs pulse duration. For both wavelengths, the very low energy components could correspond to an autoionization process. On a timescale of 8 ps, the population created in $S_4$ has completely relaxed to $S_2$. Consequently, photoionization to $D_0$ with a $\Delta v \sim 0$ propensity rule is the only signature observed in the photoelectron spectrum. In contrast, when the experiment is performed on the fs scale, a significant contribution of the photoelectron signal arises from ionization of the $S_4$ state, since photoioization takes place on a timescale commensurate with internal conversion. These observations set an upper limit certainly lower than 8 ps for the $S_3$ and $S_4$ states lifetimes. The two-colour experiment discussed in part C will fix a new upper limit to 120 fs. In conclusion, photoionization can be driven simply by $\Delta v \approx 0$ transitions to $D_1$ from the $S_4$ state and to $D_0$ from the $S_2$ state.

**C – Two Colour photoelectron spectra**

Having identified the $S_4$ contribution to the total photoionization cross-section, the non-radiative processes related to the decay times of Figure 2 might be assigned by time-resolved photoelectron spectroscopy. Note that in the two-colour experiments the probe pulse is centered around 400 nm, so it is useful to compare the $\Delta v \approx 0$ propensity rule observed in 1+1 photoionization (at 266 nm) to the 1+1' or 1+2' schemes :

*One colour photoionization 1 + 1*

$$S_0 \xrightarrow[266nm]{h\nu} S_4(\text{v}=0) + E_{vib}(0.26\ eV) \xrightarrow[266nm]{h\nu} D_1(\text{v}=0) + 0.82\ eV \,|\Rightarrow \Delta\text{v}=0 \text{ allowed}$$

$$D_0(\text{v}=0) + 1.92\ eV \,|\Rightarrow \Delta\text{v}=0 \text{ allowed}$$

$$\downarrow IC$$

$$S_2(\text{v}=0) + E_{vib}(1.1\ eV) \xrightarrow[266nm]{h\nu} D_o(\text{v}=0) + 1.92\ eV \,|\Rightarrow \Delta\text{v}=0 \text{ allowed}$$

*Two colour photoionization 1 + 1′*

$$S_0 \xrightarrow[266nm]{h\nu} S_4(\text{v}=0) + E_{vib}(0.26\ eV) \xrightarrow[400nm]{h\nu} D_1(\text{v}=0)\text{-}0.74\ eV \,|\Rightarrow \text{ionization not allowed}$$

$$D_0(\text{v}=0) + 0.35\ eV \,|\Rightarrow \Delta\text{v}=0 \text{ allowed}$$

$$\downarrow IC$$

$$S_2(\text{v}=0) + E_{vib}(1.1\ eV) \xrightarrow[400nm]{h\nu} D_o(\text{v}=0) + 0.35\ eV \,|\Rightarrow \Delta\text{v}=0 \text{ not allowed}$$

$$S_0 \xrightarrow[335nm]{h\nu} S_2(\text{v}=0) + E_{vib}(0.14\ eV) \xrightarrow[400nm]{h\nu} D_o(\text{v}=0) - 0.61\ eV \,|\Rightarrow \text{ionization not allowed}$$

*Two colour photoionization 1 + 2′*

$$S_0 \xrightarrow[266nm]{h\nu} S_4(\text{v}=0) + E_{vib}(0.26\ eV) \xrightarrow[400nm]{2h\nu} D_1(\text{v}=0) + 2.36\ eV \,|\, \Delta\text{v}=0 \text{ allowed}$$

$$\downarrow IC$$

$$S_2(\text{v}=0) + E_{vib}(1.1\ eV) \xrightarrow[400nm]{2h\nu} D_o(\text{v}=0) + 3.45\ eV \,|\Rightarrow \Delta\text{v}=0 \text{ allowed}$$

$$S_0 \xrightarrow[335nm]{h\nu} S_2(\text{v}=0) + E_{vib}(0.14\ eV) \xrightarrow[400nm]{2h\nu} D_o(\text{v}=0) + 2.49\ eV \,|\Rightarrow \Delta\text{v}=0 \text{ allowed}$$

With a 400 nm probe 1+1’ photoionization onto $D_0$ is energetically only allowed for pump wavelengths shorter than 287.5 nm. Photoionization onto $D_1$ or $D_2$ requires two probe photons for all the pump wavelengths studied. Taking into account the pump wavelength range and the resulting vibrational energy in the $S_2$ state, one-photon ionization at 400 nm can never satisfy the $\Delta\text{v} \approx 0$ propensity rule. Consequently, two-photon ionization is expected to compete very efficiently with one-photon ionization.

Figure 4a shows the photoelectron spectra recorded at a 1 ps delay for various pump wavelengths. Surprisingly, the photoelectron spectrum does not depend on the excitation wavelength; only the integrated photoelectron intensity varies as a function of the delay time. The very slow photoelectron (<200 meV) contribution assigned to the 1+1' route is significantly reduced at 335 nm where photoionization can only take place via a 1+2' scheme. As expected, photoionization is strongly dominated by the 2-photon transition. This is further supported by the laser power dependence of the parent ion signal, which was found to be linear in the pump intensity and quadratic in the probe intensity. If the same $\Delta v \approx 0$ propensity rule as observed in Fig. 3 is applied for two-photon ionization at 400 nm, the photoelectron correlated with $D_1$ from $S_4$ on excitation at 266 nm should appear at a kinetic energy around 2.1 eV. There is no component at any delay that appears around this kinetic energy. Indeed, the most striking feature is the similarity between the photoelectron spectra for excitation to various internal energies in $S_4$ ($\lambda_{pump}$ = 268 nm, 275.5 nm and 283 nm) and the one recorded from the $S_2$ ($\lambda_{pump}$ =335 nm) state. The invariance of the photoelectron spectra to excitation energy and pump-probe delay can only be rationalized if the two-photon probe ionization occurs out of the same electronic state for *all* of the pump energies studied. The obvious candidate is the $S_2$ state since it is the only state to have a sufficiently long radiative lifetime to be compatible with the long decay times shown in Fig. 2.

Our time resolution is evidently not sufficient to detect the transition between ionization from $S_4$ and from $S_2$ via any variation of the photoelectron spectrum, so we can conclude that the internal conversion between $S_4$ and $S_2$ is faster than 120 fs. Once in the $S_2$ state the population can convert further to highly vibrationally excited levels in $S_1$ and $S_0$. We will examine this internal conversion, which occurs over a timescale of more than 60 ps in section D. The two-colour photoionization takes place via path (d) and/or (e) of Figure 1.

As we have noted, apart from its relative intensity, the photoelectron energy spectrum is invariant to the pump photon excitation energy. The photoelectron spectra in Figure 4a are also highly structured. At first sight, it is tempting to identify this structure as a signature of the vibrational structure of the ion but this hypothesis must be immediately rejected. Indeed this pattern spreads over more than 1.5 eV with a splitting varying around 500 meV. Given the large variation of vibrational energy in $S_2$ (see Eq. 1) as well as the typical vibrational quanta of azulene cation (smaller than 0.2 eV),[23] it is difficult to assign the photoelectron pattern of Figure 4a to a vibrational progression. In fact, we will now show that this photoelectron spectrum is the signature of two-photon transitions via vibrationally excited Rydberg states.

Consider a Rydberg state lying at an electronic energy $R_n$, reached after absorption of one probe photon. Energy conservation gives

$$\hbar\omega_{\text{pump}} + \hbar\omega_{\text{probe}} = R_n + E_{\text{vib}}^{R_n} \qquad (1)$$

where $E_{\text{vib}}^{\text{R}}$ is the vibrational energy in the Rydberg state. If the molecule absorbs a second probe photon and ionizes, the ejected photoelectron will have a kinetic energy given by:[24]

$$E_{\text{kin}}^{e} = \left(\hbar\omega_{\text{pump}} + \hbar\omega_{\text{probe}}\right) + \hbar\omega_{\text{probe}} - E_{\text{vib}}^{\text{ion}} - IP \qquad (2)$$

Introducing the binding energy $E_b^n = IP - R_n$ of the Rydberg state relative to the $IP$ to which it converges, and assuming the propensity rule $\Delta\text{v} \approx 0$ in the ionization step (on Franck-Condon arguments between the Rydberg states and the cation) so that $E_{\text{vib}}^{R_n} \approx E_{\text{vib}}^{\text{ion}}$, we get:

$$E_{\text{kin}}^{e} = \hbar\omega_{\text{probe}} - E_b^n \qquad (3)$$

which demonstrates that, within such an ionization scheme, the kinetic energy distribution of the photoelectron spectrum does not depend on the pump wavelength but only on the Rydberg state binding energy and on the probe wavelength. This invariance is mainly due to the conservation of vibrational energy in the Rydberg states through the photoionization, as represented on Figure 1.2.

Such behaviour has been observed previously in the phototoelectron spectra of other aromatic systems obtained with short optical pulses.[25,26] The bandwidth of the observed photoelectron bands is relatively narrow (around 200 meV) compared to the few eV of vibrational excitation in these Rydberg states. This is another signature of photoionization taking place via a vibrational propensity rule $\Delta\text{v} \approx 0$, and further supports the assignment of the intermediate levels to Rydberg states in the resonantly induced 2-photon ionization. The binding energy with respect to $D_0$ are used to define the top axis of Figure 4a. These Rydberg states have previously been observed by Weber *et al.* via one colour 3-photon ionization with femtosecond pulses at 265 nm (MPI).[27] In this paper, the authors suggest that the Rydberg states are optically dark states populated by ultrafast internal conversion from an optically bright superexcited state (SES) of short lifetime.[28,29] A few of the Rydberg states of azulene have been detected directly by absorption spectroscopy[30,31] and by multiphoton ionization.[26,27] In the MPI studies by Weber *et al.*,[26,27] no assumption was made as to which limit the Rydberg states were converging to. Therefore, in order to compare our results to these previous data, the Rydberg states are listed in Table 1 as a function of their binding energies; their absolute electronic energies are obtained by assuming a common convergence limit to $D_0$. Weber *et al.* deduce the electronic symmetry of the Rydberg states on the basis of the quantum defect value extracted from their analysis (close to 1 for a *s* orbitals, 0.3-0.5 for *p* orbitals, and <0.1 for higher angular momentum states). Using the same arguments, the calculated quantum defects are listed on top of each contribution in Figure 4a. The binding energies extracted from Figure 4a are in good agreement with the known spectroscopy of the Rydberg states in azulene.[26,27,30,31] Based on the relative agreement with the absorption spectroscopy by Lewis *et al.*,[31] we cautiously conclude that most of the Rydberg states observed in Fig. 4a converge onto $D_0$, but this statement is amenable to further confirmation.

The remaining question is: does the photoionization from $S_2$ take place via an optical resonance with the Rydberg states (path (d) in Fig. 1) or via some other state that relaxes to Rydberg states (path (e)) as proposed by Weber? The general concept of a "superexcited state" defines a neutral state higher than the ionization potential with two-electron holes or an inner-shell electron.[32] Detailed experimental and theoretical data on the ionization of azulene are scarce,[21,22,33,34] but the onset of shake-up ion states has been calculated by Deleuze to lie at quite low excitation energies, around 9.6 eV with a dominant configuration corresponding to $...(1b_1)^2(2b_1)^2(1a_2)^2(3b_1)^2(2a_2)^0(4b_1)^0(3a_2)^1$.[33] Any neutral state converging to this cation state is likely to be a SES. The presence of a doubly excited state lying at an energy as low as 4.95 eV has been detected by magnetic circular dichroism (MCD) on different derivatives of azulene.[24] This kind of state is not strictly speaking a SES since it lies below the IP, but some of the properties of such a state are very similar to a SES. Now we will demonstrate why the photoelectron spectra of Fig. 4a lead us to propose a photoionization scheme corresponding to path (e) in Fig. 1.

If a progression of the Rydberg states involved in the time-resolved photoionization scheme (presented in Fig. 4a) converge to $D_0$, they will share the geometry of the cationic state. Consequently, transitions from $S_2$ to this subset of Rydberg states should also follow the $\Delta v \approx 0$ propensity, since $S_2$ is also parallel to $D_0$.[17,21] At a fixed probe energy, such a propensity rule might be satisfied for one of the Rydberg states but it is very unlikely to be satisfied for all of them over the implemented pump energy range. The invariance of the Rydberg features observed in the photoelectron spectra to the pump wavelength therefore implies that $D_0$ Rydberg states must be populated via an intermediate state with a different geometry to $S_2$ and $D_0$. Furthermore this intermediate state must couple to such a $D_0$ Rydberg progression via internal conversion on the timescale of the probe pulse. Once populated the Rydberg states are photoionized by a subsequent probe photon. Since the combined energy of one pump and one probe photon for $\lambda_{pump}$>287.5 nm is below the ionization threshold to $D_0$, the excited state in question is most likely an unstable double hole state with very diffuse electronic orbitals. The natural decay should then occur by an Auger like process, namely by filling the unoccupied lower molecular orbital with the resulting excess energy balanced by the ejection of an electron. Since this excited state lies below the ionization threshold, the energy required to expel the electron is not available. Nevertheless, states with an appropriate electronic configuration and an equivalent energy are present, the states being the Rydberg series converging to $D_0$. The instability of a doubly excited state would result in efficient and extremely fast internal conversion with respect to the probe pulse duration. Note that this process would be independent of the vibrational energy in this double hole state, and this would rationalize the independence of the Rydberg fingerprint to the vibrational energy in the $S_2$ manifold. This two-photon ionization channel (path(e)) would also be strongly favoured because of inefficient Franck-Condon factors from $S_2$ onto $D_0$ (path(d)) as a result of insufficient energy to achieve vibrational energy conservation (as shown in Eq. 1). Another interesting feature in the photoelectron spectra (Fig. 4a) is the prominent contribution of the Rydberg states with larger binding energy whatever the vibrational energy of the $S_2$ state. This unexpected invariance over the studied 1 eV pump energy range,[28,35] might be rationalized by an intermediate excited state originating at low electronic energy but populated with a large vibrational energy. This would be consistent with the doubly excited state revealed in MCD around 4.95 eV.[24] In such a case, internal conversion would be expected to favour population transfer to Rydberg states with a large binding energy. We can tentatively assign the electronic configuration of this doubly excited state. The $S_2$ state is a superposition of two one electron excitations: HOMO to HOMO+2 and HOMO-1 to LUMO[36]. The ion shake-up state calculated by Deleuze[33] corresponds to the excitation HOMO to HOMO+2. The doubly excited state could easily involve the double excitation of the HOMO+2 orbital. This configuration would be in agreement with Deleuze's calculation and be of $A_1$ symmetry, as required for the observed optical anisotropy[11].

Photoelectron spectra 5 and 6 of Figure 4a are one colour profiles recorded at 400 nm (5) in the femtosecond regime or (6) nanosecond regime. Remarkably, the intensity distribution of the photoelectron spectra at 400 nm is the same regardless of whether the MPI takes place in the nanosecond or femtosecond regime. It is useful here to comment on the difference between one colour photoionization at 400 nm (Figure 4a) and at 266 nm (Figure 4b). Photoelectron spectra recorded at 400 nm show the Rydberg fingerprint whilst those recorded at 266 nm display the electronic character of the $S_4$ state and the $S_2$ states. A two-photon excitation at 400 nm can reach an excited state, as already observed in the liquid phase,[3] whose population would then decay rapidly to the $S_2$ state. In that case photoionization at 266 nm via $\Delta v \approx 0$ transitions are energetically allowed, whilst these channels are closed for photoionization at 400 nm:

$$S_0 \xrightarrow[266nm]{h\nu} S_4\left(\text{v}=0\right)+E_{vib}\left(0.26\ eV\right) \xrightarrow[266nm]{h\nu} D_1\left(\text{v}=0\right)+0.82\ eV\Big| \Rightarrow \Delta\text{v}=0 \text{ allowed}$$

$$\downarrow IC$$

$$S_2\left(\text{v}=0\right)+E_{vib}\left(1.1\ eV\right) \xrightarrow[266nm]{h\nu} D_o\left(\text{v}=0\right)+1.92\ eV\Big| \Rightarrow \Delta\text{v}=0 \text{ allowed}$$

$$S_0 \xrightarrow[400nm]{2h\nu} S_{??}$$

$$\downarrow IC$$

$$S_2\left(\text{v}=0\right)+E_{vib}\left(2.64\ eV\right) \xrightarrow[400nm]{h\nu} D_o\left(\text{v}=0\right)+1.92\ eV\Big| \Rightarrow \Delta\text{v}=0 \text{ not allowed}$$

Consequently at 400 nm the transition to the intermediate double hole state competes efficiently with direct photoionization, in contrast to the 266 nm scheme.

Doubly excited states are quite difficult to investigate spectroscopically mainly due to their markedly different geometry to the valence states as well as their short lifetimes. It is mainly through their relaxation process that they are detectable. Despite the difficulty, we hope our observations might inspire theorists to take up the challenge to calculate the geometry and electronic configuration of the doubly excited states on polyaromatic systems, such as azulene. Indeed, these curious states are of potential importance in a variety of interesting processes since these states might coexist with the ion states as well as dissociative neutral states. The dynamics of such states are thus likely to play a crucial role in Rydberg fingerprint spectroscopy, used for instance to differentiate charge distributions in isomeric aromatic molecules,[35] as well as conformational dynamics.[37]

**D - Non-radiative relaxation from vibrationally excited $S_2$ state**

The time transients recorded in Figure 2 map the relaxation from the $S_2$ state. This is revealed by the persistence of the intensity distribution of the time-resolved photoelectron spectrum as a function of the pump-probe delay as well as the same sensitivity to the rotational coherence (not presented here). We now examine the variation of the decay rate as a function of the excitation energy using a statistical approach.

Non-radiative relaxation in azulene is largely governed by internal conversion (IC) rather than intersystem crossing (ISC). If we ignore ISC, the decay time $\tau_e$ is simply a function of the radiative rate $k_r$ and the non-radiative relaxation rate $k_{IC}$:

$$\frac{1}{\tau_e} = k_e = \frac{k_r}{\phi_r} = k_{IC} + k_r \tag{4}$$

where $\phi_r$ is the fluorescence quantum yield. In Figure 5 we plot the decay rates measured for different vibrational energy in $S_2$ state, together with lifetime measurements[9] obtained for low $E_{vib}$ and data deduced from quantum yield measurements.[7,8] The fluorescence rate is usually supposed independent of the vibrational mode,[38] such that any variation in the decay time simply stems from a dependency of the IC rate with the excess energy. At the origin of $S_2$, the weak coupling between the $S_2$-$S_1$ state[39] can be described by an electronic energy gap law.[40] At larger excess vibrational energies the relaxation becomes dominated by the $S_2$-$S_0$ internal conversion. [7]

With increasing vibrational energy, the internal conversion rate is expected to increase slightly before tailing off at yet higher energy.[38,41] The overall trend of exponentially increasing $k_{IC}$ versus vibrational energy in $S_2$ has already been observed[7-9,41] for an excess vibrational energy greater than 0.24 eV.[2,8] In the framework of Fermi's Golden Rule, the internal conversion rate can be approximated as:[38]

$$k_{IC}(E_i) = \frac{2\pi}{\hbar} |V_i|^2 \rho(E_i) \tag{5}$$

where $\rho(E_i)$ is the vibrational level density in $S_0$ at the energy $E_i = E_{S_2}(0) + E_{vib}(S_2)$ with $E_{S_2}(0)$ defining the vibrational zero point energy of $S_2$ and $|V_i|^2$ the mean-square value of the $S_2$-$S_0$ coupling at the energy $E_i$. The vibrational level density at high excess energy can be roughly estimated by an exponential function of vibrational energy:[38]

$$\rho(E_i) \propto \exp\left(\frac{E_i}{\varepsilon_{S_0}}\right) \tag{6}$$

where $\varepsilon_{S_0}$ is the typical energy scale of increase in level density as a function of the vibrational energy in $S_0$. In a crude approximation, one can assume a weak variation of coupling $|V_i|^2$ as a function of $E_i$, such that:

$$\alpha = \frac{\delta \ln(k_e)}{\delta E_{vib}(S_2)} = \frac{k_{IC}}{k_r + k_{IC}} \frac{1}{\varepsilon_{S_0}} \sim \frac{1}{\varepsilon_{S_0}} \tag{7}$$

This simple calculation overestimates α by not taking into account the vibrational modes promoting the IC that will lead to a reduced state density, however, it explains the linear dependency observed in Figure 5.

Figure 5 compares the rate $k_e$ measured at high excess energy in the present work with earlier gas-phase measurements at lower excess vibrational energy.[7-9,41] The value of $\alpha = 1.70 \pm 0.05$ eV$^{-1}$ is the result of a linear fit taking into account our data together with two measurements obtained by picosecond resolved fluorescence and a jet-cooled molecular beam at excess energy ≥ 236 meV.[8]

In order to detect the importance of promoting modes, a calculation of the state density in $S_0$ at the different vibrational energies of $S_2$ has been carried out using the generating function for the number of states of a given energy:[42]

$$f(z) = \prod_{i=0}^{n_{A_1}} \frac{1}{1 - z^{\varepsilon_i}} \tag{8}$$

where $n_{A_1}$ is the total number of in-plane vibrational modes of $A_1$ symmetry (17 for azulene) and $\varepsilon_i$ are the vibrational energies in $S_0$.[23] The number of possible combinations at the total vibrational energy E in the $S_0$ state is calculated by numerical integration of the Cauchy residue formula which inverts the generating function:[43]

$$N(E) = \frac{1}{2\pi i} \int_c \frac{dz}{z^{E+1}} f(z) \tag{9}$$

This contour integral is done with a resolution of ΔE=10 meV, fixed by the convergence of the calculation, leading to a density of states:

$$\rho(E)=\frac{N(E)}{\Delta E} \tag{10}$$

Here the anharmonicities of the vibrational modes are not taken into account nor are all the mode combinations, but the good agreement with the experimental data (slope $\alpha = 1.21 \pm 0.02$ eV$^{-1}$ - dashed plot on Figure 5) leads us to conclude that the internal conversion $S_2$- $S_0$ is not limited by a promoting mode process. Note that, for instance, if the 6 out-of plane $A_2$ vibrational modes[23] are added in the calculation, the slope increases to $\alpha = 2.27 \pm 0.02$ eV$^{-1}$.

At high excitation energy, the internal conversion rate might simply depend on the difference in entropy between the initial and the final electronic state. Indeed, the entropy of the initial state has to be taken into account to reflect the dissipative character of the intramolecular vibrational relaxation. When the vibrational energy in $S_2$ becomes significant, the entropy of $S_2$ increases at the same rate $\varepsilon = \varepsilon_{S_0}$ as the entropy in $S_0$ and the internal conversion rate will reach its converging limit defined by:[38]

$$k_{IC} = \frac{2\pi}{\hbar}\frac{|V_i|^2}{\varepsilon}\exp\left(\frac{E_{S_2}(0)-E_{S_0}(0)}{\varepsilon}\right) \tag{11}$$

Although we have not yet observed this fall-off behaviour in the IC rate between the $S_2$ and $S_0$ states, we can still estimate a lower limit for the vibronic coupling $V_i$ from the $k_{IC}$ measured at the maximum excess vibrational energy investigated here (1.24 eV) s$^{-1}$ and $\varepsilon \propto \frac{1}{\alpha} = 586 \pm 17$ meV. The lower limit for $V_i$ is 60 ± 5 μeV. Note that the $S_2$-$S_4$ vibronic coupling is estimated to 99.2 meV.[20] A more definitive statement concerning the magnitude of the coupling must await the determination of the fall-off region of $k_{IC}$ via pump-probe experiments done at higher energy. $\alpha = 1.50 \pm 0.53$ eV$^{-1}$.

## IV CONCLUSION

We have performed time resolved studies on the $S_2$-$S_4$ excited states of azulene. Our pump-probe experiments reveal an internal conversion process onto the $S_2$ state, followed by an internal conversion onto $S_0$. The $S_2$-$S_0$ decay rate follows an exponential energy gap law as a function of the vibrational energy in the $S_2$ state. The vibronic coupling strength is anticipated to be larger than 60 μeV based on a statistical description of the internal conversion. These experiments also reveal an uncommon ionization pathway involving a common set of Rydberg states that are populated on an ultrafast timescale from a doubly excited state, or possibly several such states. Experimental evidence suggests that this probe transition becomes efficient when the excess energy in the ion continuum is not high enough to reach the Franck-Condon window. Above the IP doubly excited states can autoionize or undergo dissociative ionization in addition to relaxing by an Auger-like process to vibrationally excited Rydberg whose origins lie below the IP. However, if the doubly excited state itself lies below the IP only this last decay route is open. In fact, these processes may be quite general in aromatic systems in which doubly excited states below the IP are likely to be ubiquitous.

### ACKNOWLEDGEMENTS

The authors would like to thank D.S. Dean for his help in calculating the density of states, and A. Beswick for helpful references on statistical theories applied to relaxation dynamics. This work was supported by CNRS, le Ministère de la Recherche, Région Midi-Pyrénées through ”Plan état-Région Spectroscopies Optiques Ultimes”, ANR COCOMOUV and the British Council Alliance programme. BJW thanks the CNRS and the UPS for his invited positions in the LCAR. KR thanks the CNRS for his postdoc fellowship. GT thanks the European network COCOMO for his postdoc fellowship. IW and BJW are also grateful to the EPSRC for partial support of this work through GR/S47656/01 and for a research studentship.

## TABLES

Table 1 : Band origins of the main electronic states of azulene and its cation. The orientation of the dipolar momenta of transition from $S_0$ are added in parenthesis. The Rydberg states are listed as a function of their binding energies and compared to various listed in the literature. All energies are in eV.

| | Electronic Energy | Electronic Energy If converging onto $D_0$ | Binding energy | Binding energy converging onto $D_0$ Ref 30 | Binding energy converging onto $D_0$ Ref 31 | Binding energy Ref 26 | Binding energy Ref 27 |
|---|---|---|---|---|---|---|---|
| $S_1$ (y) | 1.77 | | | | | | |
| $S_2$ (z) | 3.56 | | | | | | |
| $S_3$ (y) | 4.23 | | | | | | |
| $S_4$ (z) | 4.40 | | | | | | |
| $R_A$ | | 4.72 | 2.69 | | 2.70 ($3p_x$) | | |
| $R_B$ | | 5.19 | 2.22 | | 2.51 ($3p_y$) | | 2.16 (3p) |
| $R_C$ | | 5.64 | 1.77 | | 1.63 ($3d_{xy}$) | | 1.62 (3d) |
| $R_D$ | | 5.92 | 1.49 | 1.40 (4s) | 1.42 ($3d_{xz}$) | | 1.36 (4s) |
| R | | 6.15 | 1.26 | | | 1.16 (4p) | 1.14 (4p) |
| R | | 6.33 | 1.08 | 1.03 (4p) | | 1.06 (4d) | 1.06 (4d) |
| $D_0$ | 7.41 | | | | | | |
| $D_1$ | 8.50 | | | | | | |
| $D_2$ | 10.07 | | | | | | |
| $D_3$ | 10.85 | | | | | | |

## FIGURE CAPTIONS

**Figure 1 :**

(1) One-colour photoionization onto the cation states $D_0$ and $D_1$ via one photon resonance with $S_4$ (path a). Within the pulse duration, internal conversion to the $S_2$ state ($IC_1$) can take place (path b). $\varepsilon_0^{(a)}$ ($\varepsilon_0^{(b)}$) is the photoelectron kinetic energy expected from photoionization taking place onto $D_0$ from the path (a) (respectively from the path (b)), assuming approximate conservation of the vibrational energy. See the Fig 3 for the recorded photoelectron spectra.

(2) And (3) Two-colour Pump-probe (1+1' or 1+2') photoionization with (c) a direct one-photon probe transition from the initially prepared state ($S_4$ in this example) or (d) a two-photon photoionization enhanced via resonances with Rydberg states $R_1$ and $R_2$. During the pump-probe delay internal conversion onto the $S_2$ ($IC_1$) and $S_0$ ($IC_2$) states can take place. (e) Two-photon photoionisation via a doubly excited state (**) that decays within the probe pulse duration onto Rydberg states ($IC_3$).

**Figure 2 :**

Time-transients recorded on the parent ion $C_{10}H_8^+$ with $\Delta t=0$ determined by a cross-correlation recorded by photionization of NO. A single exponential fit (full line) gives a time constant of (a) 273.8+/-10.5 ps for an excitation taking place at the origin band of the $S_3$ state, namely $\lambda_{pump}$=293 nm and (b) 68+/-1 ps at 223 meV above the origin band of the $S_4$ state, namely $\lambda_{pump}$=268 nm. For both measurement, the probe pulse is the second harmonic generation of the fundamental output (~400 nm).

**Figure 3 :**

One-colour photoelectron spectra recorded with a fs-pulse centered at 266 nm (empty circle) or 258 nm (full square). For both wavelengths, vertical lines indicate the maximum kinetic energy that might be released for ionization taking place to the cation ground state $D_0$ or the first excited state $D_1$.

**Figure 4 :**

Photoelectron spectra recorded with a probe pulse centered at (a) 400 nm and pump excitation at (1) 268 nm (2) 275.5 nm, (3) 283 nm (4) 335 nm. For comparison one-colour photoionization at 400 nm obtained (5) in the femtosecond regime or (6) nanosecond regime are plotted, as well as (b) the 266 nm fs-one colour experiment presented in Fig 3. Each 1+2' photoelectron spectra has been recorded at $\Delta t \sim 1$ps after subtraction of the background spectra (one-colour experiment recorded with the pump alone or the probe alone).

The vertical shifts introduced to compare the spectra are indicated by horizontal dot lines. The different Rydberg states are identified as a function of their binding energy (with a probe step at one-photon of 400 nm-top axis of (a)), their principal quantum number n and a quantum defect $\delta$ assuming that these ones are in the 0-1 range.

**Figure 5 :**

Plot of $\log(k_E)$ versus vibrational energy in the $S_2$ state. Our data are compared to previous work by Hirata et al. [7], Abou-Zield et al [9] and Demmer et al [8], as well as to the calculated density of states taking into account only the vibrational modes of $A_1$ symmetry. Linear fits are done on our data (straight line) and the calculated density of states (dashed line).

V. Blanchet et al., Fig. 1

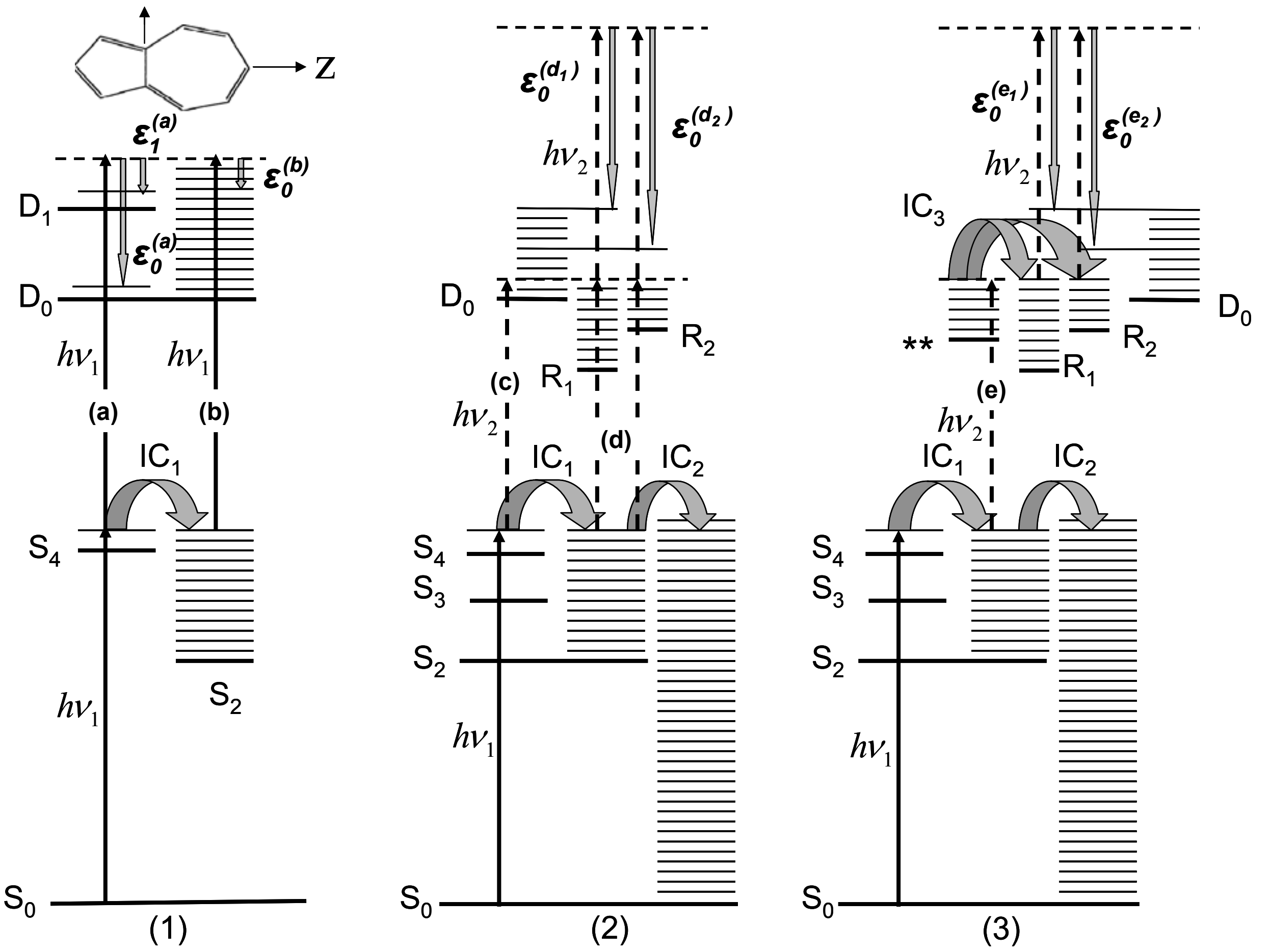

V. Blanchet et al., Fig. 2

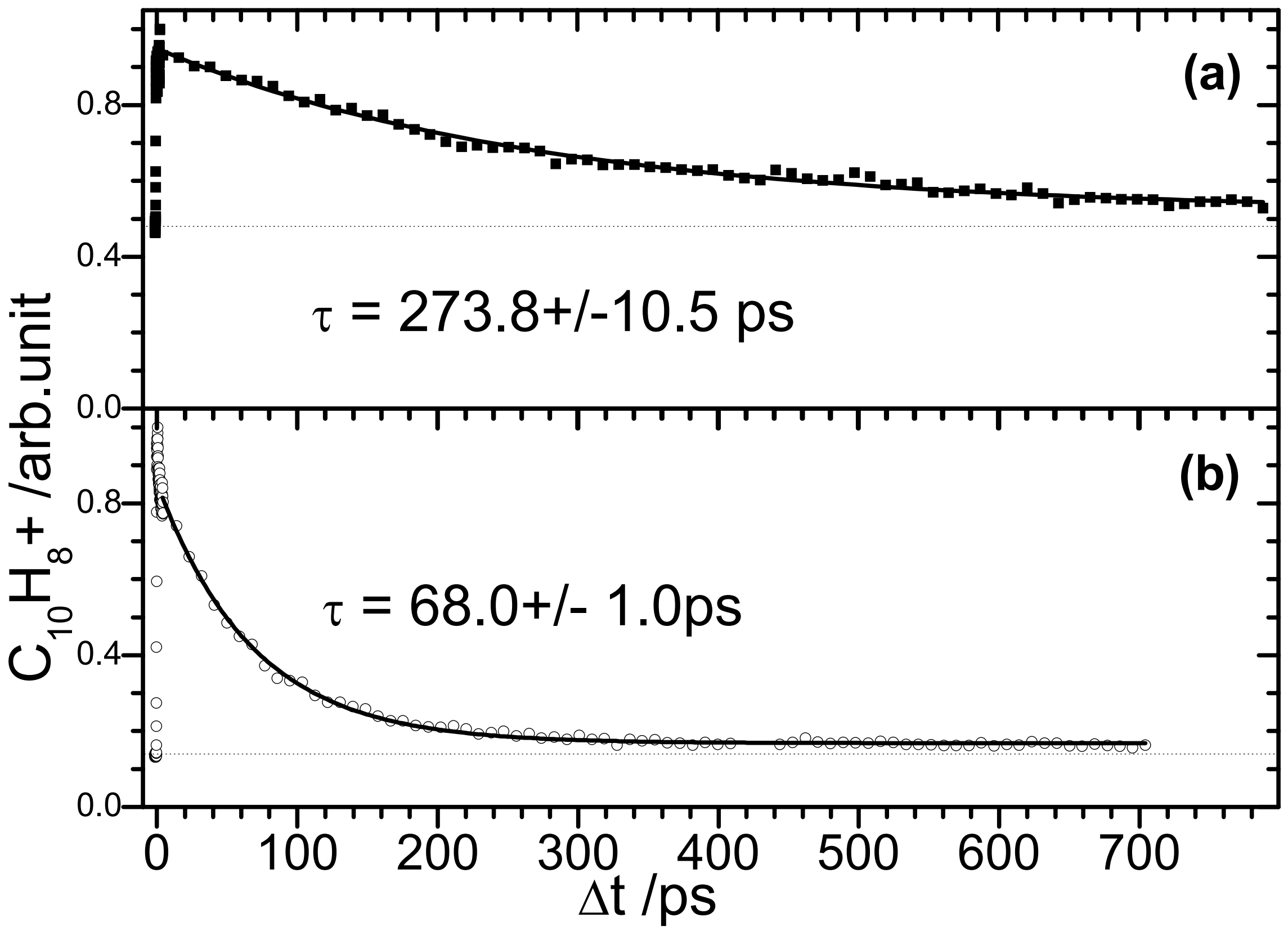

V. Blanchet et al., Fig. 3

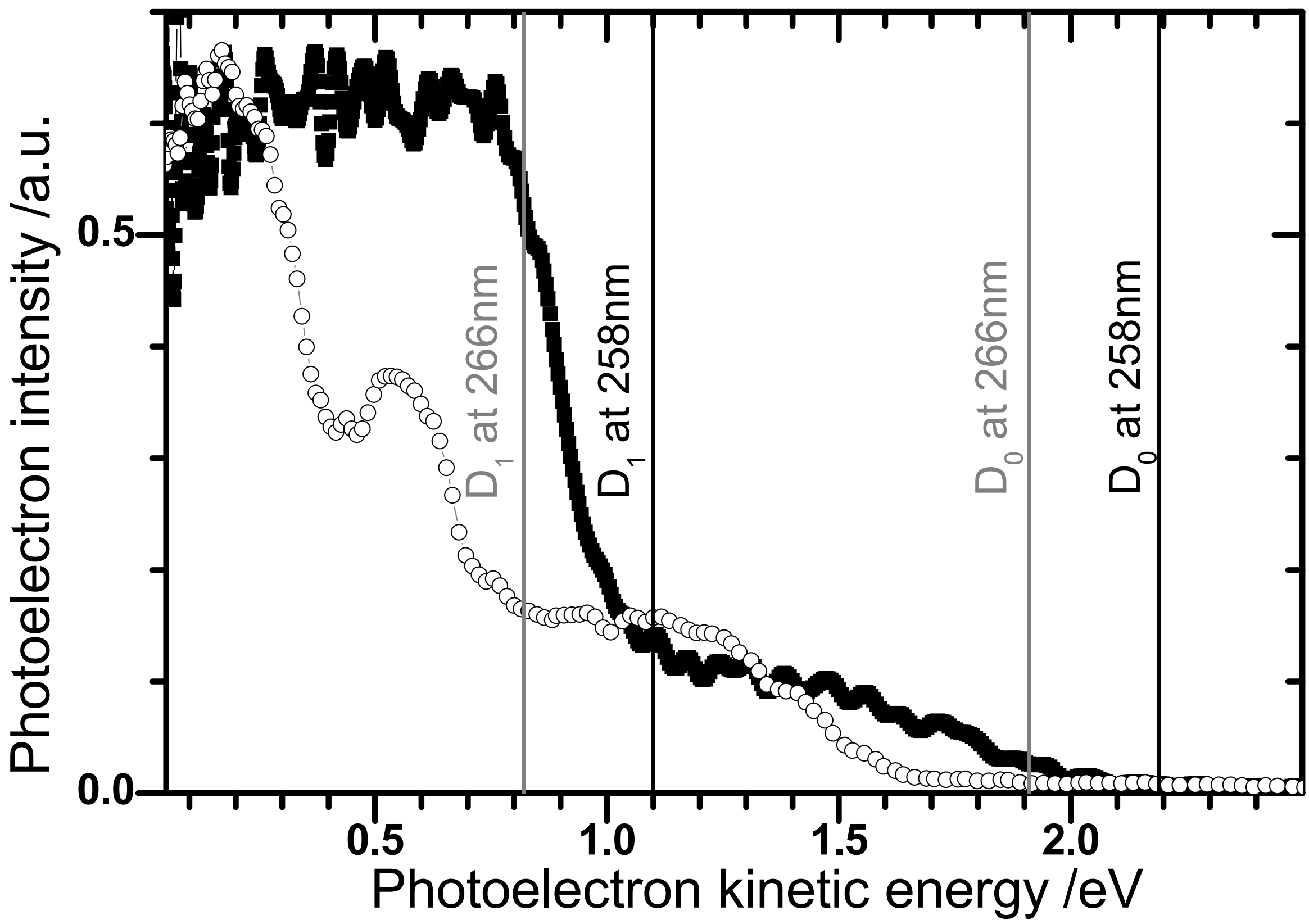

V. Blanchet et al., Fig. 4

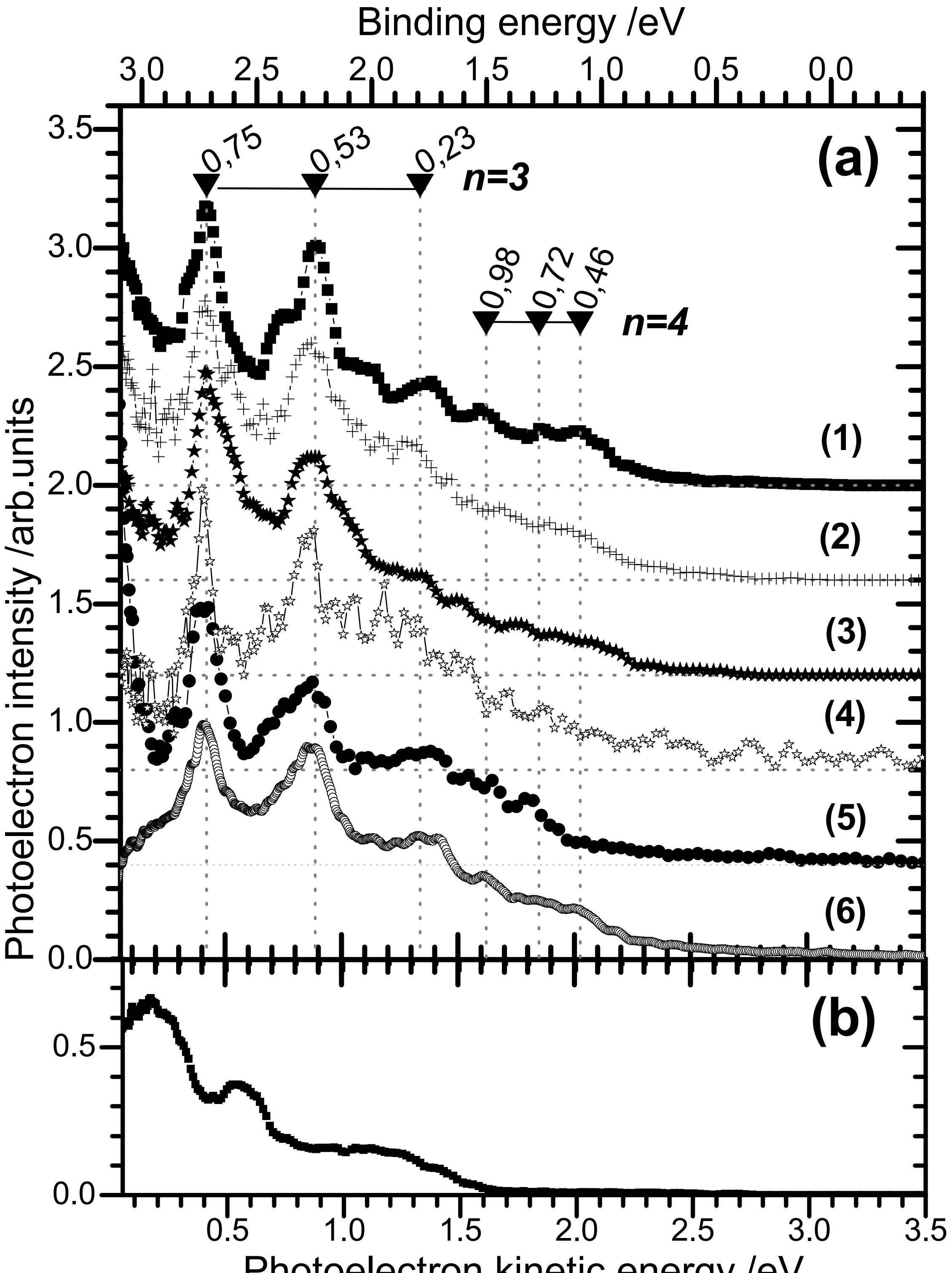

V. Blanchet et al., Figure 5

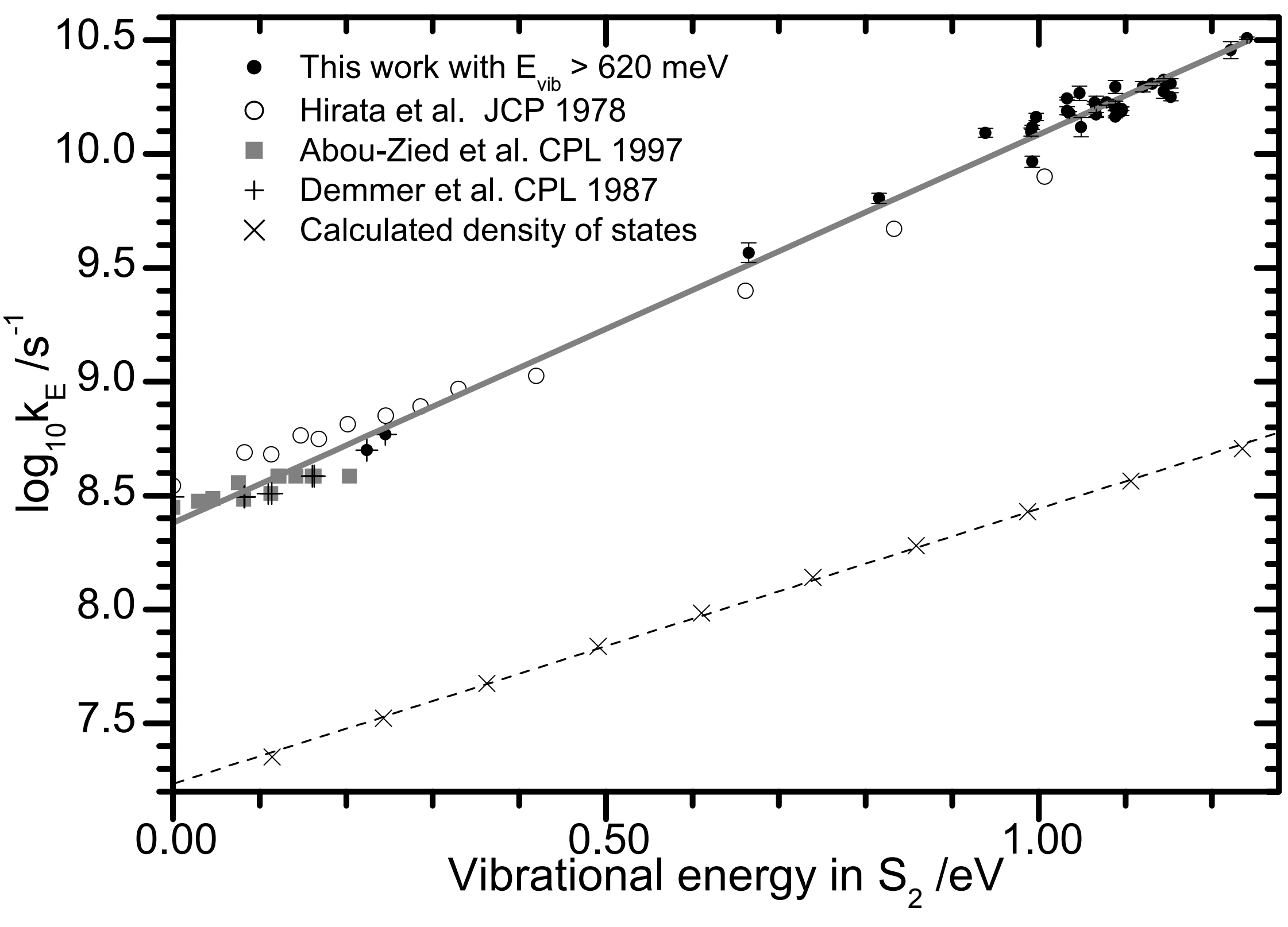